\documentclass[twocolumn,preprintnumbers,secnumarabic,amsmath,amssymb,superscriptaddress,nofootinbib]{revtex4}

\usepackage{graphicx}
\usepackage{xcolor}
\usepackage{dcolumn}
\usepackage{bm}

\usepackage{tikz}
\usetikzlibrary{3d,calc}
\usepackage{amssymb,amsmath}
\usepackage{bm} 
\usepackage{booktabs} 
\usepackage{array}
\usepackage{latexsym}
\usepackage{graphicx}
\usepackage{color}
\usepackage{slashed}
\usepackage{datetime}
\usepackage{verbatim}
\usepackage{enumerate}
\usepackage{url}
\usepackage{chngpage} 
\allowdisplaybreaks
\usepackage{psfrag}
\usepackage{mathtools}
\usepackage{mciteplus}
\usepackage{dsfont}
\usepackage{multirow} 
\usepackage{bbm} 
\usepackage[justification=justified, singlelinecheck=false]{caption}

\usepackage{subcaption}
\usepackage{stmaryrd}
\usepackage[colorlinks=true,      linkcolor=blue,      urlcolor=blue,      
            filecolor=blue,      citecolor=blue,       pdfstartview=FitH,     
						pdfpagemode=UseNone,      bookmarksopen=true]{hyperref}  
\usepackage[all]{hypcap}     

\usepackage{tikz}
\usepackage{pgfplots}
\usepackage{amsmath}
\pgfplotsset{compat=1.16}
\usepackage{amsmath}

\definecolor{cardinal}{rgb}{0.6,0,0}
\definecolor{darkgreen}{rgb}{0,0.4,0}
\definecolor{golden}{rgb}{0.92, 0.7, 0}
\definecolor{midnight}{rgb}{0, 0, 0.5}
\definecolor{darkblue}{rgb}{0, 0, 0.7}
\definecolor{purple}{rgb}{0.5, 0, 0.5}

\def\ZZ{\mathbb{Z}}

\def\IR{\mathbb{R}}

\newcommand{\0}{\textrm {\tiny{(0)}}}

\newcommand{\beq}{\begin{equation}}
\newcommand{\eeq}{\end{equation}}

\newcommand{\bea}{\begin{eqnarray}}
\newcommand{\eea}{\end{eqnarray}}

\def\ZZ{\mathbb{Z}}

\begin{document}

\vspace*{-.1cm}
\title{Lost in Translation: Moduli Stabilization from EFT to Eleven Dimensions}

\smallskip
\author{Iosif Bena}

\affiliation{{Institut de Physique Th\'eorique,
	Universit\'e Paris Saclay, CEA, CNRS,
 F-91191 Gif-sur-Yvette, France
}}

\author{Rapha\"el Dulac}

\affiliation{{Institut de Physique Th\'eorique,
	Universit\'e Paris Saclay, CEA, CNRS,
 F-91191 Gif-sur-Yvette, France
}}

\author{Mariana Gra\~na}

\affiliation{{Institut de Physique Th\'eorique,
	Universit\'e Paris Saclay, CEA, CNRS,
 F-91191 Gif-sur-Yvette, France
}}

\author{Dimitrios Toulikas}

\affiliation{{Department of Physics,
Ben-Gurion University of the Negev,
Beer Sheva 84105, Israel
}}

\preprint{}

\begin{abstract}

We explicitly show how moduli stabilization is realized geometrically in M-theory compactified on $T^4/\mathbb{Z}_2\,  \times\,  $K3, by using the Gibbons-Hawking approximation of the K3 metric. By relating this compactification to certain microstate geometries, we present the explicit solutions in which fully backreacted fluxes on certain four-cycles stabilize three of the $T^4/\mathbb{Z}_2$ compactification moduli. The minimal tadpole contribution of these fluxes is linear  in the number of stabilized moduli, and we argue that this linear relation holds for more general fluxes. 

We also construct a one-parameter family of supersymmetric eleven-dimensional solutions that break Lorentz invariance and the warped-product structure of the compactification. These solutions are a continuous deformation of the warped-product Lorentz-invariant compactification, to which they reduce  when the moduli reach their stabilized values. Away from the Lorentz-invariant locus, the fluxes are no longer self-dual in the internal space, and include fields that do not exist in the corresponding EFT. Remarkably, although these fluxes still stabilize a modulus, it is not the $T^4/\mathbb{Z}_2$ modulus that appears stabilized in the Lorentz-invariant solution, but rather a nontrivial combination of the $T^4$ volume and K3 shape moduli.

The existence of these solutions suggests that the EFT description of moduli stabilization can be misleading and does not reflect the moduli-stabilization dynamics of the full eleven-dimensional theory.
Our results extend straightforwardly to Type IIB String Theory compactified on orientifolds of $T^2 \times K3$.

\end{abstract}

\maketitle

\section{Introduction}
\vspace*{-.3cm}

Generic compactifications of string theory to four dimensions give rise to a large number of massless scalar fields (moduli) which parametrize the shape and size of the internal space. Since no such fields are observed in nature, they must be made massive in any realistic compactification. Fluxes threading the internal space generate a scalar potential for some of these moduli, providing a perturbative stabilization mechanism. In type IIB warped compactifications on Calabi-Yau three-folds, this potential descends from the Gukov-Vafa-Witten superpotential \cite{Gukov:1999ya}, which stabilizes the complex structure moduli. An analogous classical mechanism arises in the three-dimensional effective theories arising from M-theory compactifications on warped Calabi-Yau four-folds \cite{Becker:1996gj}, where fluxes similarly stabilize a subset of the moduli.

However, despite much progress in understanding the classical stabilization mechanism of complex structure moduli at the level of the three- and four-dimensional low-energy effective theories \cite{Kachru:2003aw,Grana:2005jc,Balasubramanian:2005zx,Douglas:2006es,McAllister:2023vgy,Marchesano:2021gyv,Lust:2025qxe}, the bulk of this effort has remained at the EFT level, and comparatively little attention has been paid to understanding how this stabilization is realized directly in string theory or M-theory. There are serious reasons for this: the metric of generic compact CY manifolds is not known, and only recently have machine-learning methods began to give us explicit expressions  \cite{Anderson:2020hux,Berglund:2022gvm,Douglas:2020hpv,Erbin:2020tks,Larfors:2021pbb,Ruehle:2020jrk} (for a very recent review see \cite{Berglund:2026hdh}). Furthermore, adding fluxes on these metrics would be much more challenging, requiring possibly extra layers on ML optimization.

The purpose of this Letter is to move beyond the low-energy-effective description of moduli stabilization, and show how it takes place directly in 11 dimensions. We explicitly construct the metric and fluxes that stabilize some of the moduli by using the fact that the  $\IR^{1,6} \times$ K3 geometry comes from uplifting to M-theory sixteen D6 branes with $\IR^{1,6}$ worldvolume, located at certain points in a $T^3/\ZZ_2$ orbifold that has eight O6 planes. The M-theory uplift of a single D6 brane is the Taub-NUT metric, and the uplift of a single O6 plane is the Atiyah-Hitchin metric, but the uplift of a combination of O6 planes and D6 branes is not known \cite{Sen:1997kz,Seiberg:1996bs,Atiyah:1985dv,Gibbons:1986df}. However, far from the O6 planes, the Atiyah-Hitchin metric resembles a Taub-NUT metric sourced by a center of charge $-4$. This allows one to write an approximate metric for K3 in the region of the moduli space where the D6 branes are away from the O6 planes \cite{Seiberg:1996bs,Schulz:2012wu}. This metric is a simple ambi-polar Gibbons-Hawking metric of the type used for constructing black-hole microstate geometries \cite{Bena:2005va,Berglund:2005vb} (for reviews see \cite{Bena:2007kg,Warner:2019jll}). The motion of the D6 branes on $T^3/\ZZ_2$ parameterizes 48 of the 58 real moduli of K3, and 16 of the 22 two-cycles of K3 are the two-cycles that run between the Taub-NUT (Gibbons-Hawking) centers and between these centers and their orientifold image (see \cite{Schulz:2012wu} for a complete discussion of  moduli and cycles of K3 in this approximation).

In this Letter we  compactify four of the $\IR^{1,6}$ directions on $T^4/\mathbb{Z}_2$ and add to the resulting $\IR^{1,2} \times T^4/\mathbb{Z}_2  \times$ K3  geometry four-form fluxes that wrap two-cycles of $T^4/\mathbb{Z}_2$ as well as  two-cycles on K3 that stretch between two Taub-NUT centers. The full supergravity solution is a restricted set of the solutions of \cite{Bena:2005va,Berglund:2005vb} and is fully known. In general, these solutions are not warped product compactifications and do not preserve the Lorentz  invariance in the 2+1 dimensional spacetime. However, we will see that imposing that the resulting solutions have a warped-product  structure and Lorentz invariance  forces some of the complex-structure moduli (describing ratios of two-cycles sizes in the original  $T^4$) to have a fixed value. 
Since the stabilization mechanism we describe is {\em local}, one can easily replace $T^4/\mathbb{Z}_2 \times$ K3 with $T^4 \times $ K3, or dualize our solution to a IIB compactification on $T^2 \times $ K3 or an orbifold thereof.

Our explicit construction of the approximate solution describing the 11-dimensional flux backgrounds addresses two important points:
\begin{itemize}

\item Since the GVW superpotential is obtained by assuming a Lorentz-invariant warped-product compactification and its associated low-energy effective theory,    one can ask whether the moduli stabilization it predicts is an artifact of imposing this ansatz.

We show that it is indeed an artifact, by explicitly constructing a continuous one-parameter family of supersymmetric compactifications that have neither a warped-product structure nor $2+1$-dimensional Lorentz invariance. Away from the Lorentz-invariant locus, the fluxes no longer stabilize the $T^4/\mathbb{Z}_2$ dimensionless deformation moduli, but rather a nontrivial combination of K3 shape moduli with the $T^4/\mathbb{Z}_2$ volume modulus --- which is famously not stabilizable by fluxes in the EFT. Furthermore, the four-form fluxes on the compactification manifold are no longer anti-self-dual.  Thus, relaxing the Lorentz-invariant warped-product ansatz reveals a qualitatively different stabilization mechanism, 
invisible to the effective description.

 \item Recently, two of the authors with Bl{\aa}b{\"a}ck and L{\"u}st conjectured  that the tadpole sourced by fluxes stabilizing a given number of moduli grows  linearly with that number~\cite{Bena:2020xrh,Bena:2021wyr}. Our explicit construction confirms this linear scaling and shows explicitly how this works even for  small numbers of moduli.

\end{itemize}

\section{K3 in the Gibbons-Hawking approximation}
\label{GH-K3}

In M-theory, the uplift of multiple D6-branes is described by a multi-center Gibbons-Hawking (GH) metric
\begin{equation}
\begin{aligned}
ds_{11}^2&=ds_{6,1}^2+ds_{4,\mathbf{B}}^2 \\
&= ds_{6,1}^2+\frac{1}{V}\left(d\psi+\vec{A} \cdot d\vec{y} \right)^2+V~ d\vec{y}^{~2}\,, 
\label{eq: metric uplift D6 brane}
\end{aligned}
\end{equation}
where $V$ is a harmonic function with poles at the positions of the D6 branes on any locally-flat three-dimensional base-space with coordinates $(y^1,y^2,y^3)$, and $\vec{\nabla}V=\vec{\nabla}\wedge\vec{A}$.

Similarly, in the M-theory uplift of an O6 plane, the metric $ds_{4,\mathbf{B}}^2$ is the smooth Atiyah-Hitchin metric \cite{Atiyah:1985dv}. Away from the O6 planes this metric  can be approximated (up to exponentially small corrections \cite{Sen:1997kz}) by a Gibbons-Hawking metric in which the poles of $V$ have charge  $-4$ times the charge of the poles of the D6-brane uplift. When one approaches these negatively-charged centers, the Gibbons-Hawking approximation breaks down (as $V$ becomes negative) and one rather needs to use the full Atiyah-Hitchin metric. 

As reviewed in Appendix \ref{App:K3bubbling}, the smooth K3 surface comes from uplifting to M-theory 16 D6 branes extended along $\IR^{6,1}$ and sitting at points in a transverse   $T^3/\mathbb{Z}_2$ space, parametrized by the coordinates $\vec y$ in \eqref{eq: metric uplift D6 brane} \cite{Seiberg:1996bs,Schulz:2012wu}. This space has 8 fixed points of the $\ZZ_2$ action, corresponding to O6 planes, and for each D6 brane we also have an image D6 brane (see Figure \ref{fig:K3-GH}). When the D6 branes are away from the O6 planes, the uplift of the region near each of the O6 planes is a smooth Atiyah-Hitchin space, so the full K3 metric is smooth. When $k$ of the D6 branes (and their $k$ image D6 branes) sit on top of an O6 plane the gauge symmetry is enhanced to $SO(2k)$ and the uplifted  geometry develops a corresponding singularity \cite{Sen:1997kz}.

We thus can approximate the K3 nonsingular metric with a GH metric with a $T^3/\ZZ_2$ base space with coordinates $(y^1,y^2,y^3)$, and a harmonic function $V$ sourced by the 16 GH centers corresponding to the D6 branes, the 16 GH centers corresponding to the image D6 branes, and 8 GH centers of charge $-4$ located at the fixed points of $T^3/\mathbb{Z}_2$ :
\begin{equation}
    V=1+\sum_{p=1}^8 V_{\text{O}6,\,p}+\sum_{i=1}^{16}V_{D6,\,i}+\sum_{i=1}^{16}V_{D6',\,i}\,.
    \label{eq: harmonic function K3}
\end{equation}
The total (GH) D6 charge in the compact base space is zero, as required by tadpole consistency. Note that the $U(1)$ isometry along $\psi$ of the GH solution is an artifact of this approximation, and is not present in the full exact K3 geometry \cite{Sen:1997kz,Schulz:2012wu}. We give more details on the GH approximation of K3 in Appendix \ref{App:K3bubbling}.

Using the fact that the K3 geometry can be approximated by an ambipolar GH metric, we can compactify four of the locally-flat $\IR^{1,6}$ coordinates, $ (x^5,x^6,x^7,x^8)$, on a $T^4/\ZZ_2$
 and leave two directions, $(x^9,x^{10})$ non-compact. This will allow us to relate solutions on which one adds four-form flux with two legs on K3 (denoted by ${\bf B}$ in \eqref{eq: metric uplift D6 brane}) and two legs on $T^4/\ZZ_2$ to a subclass of the bubbling microstate geometries with an ambipolar GH base-space constructed in M-theory \cite{Bena:2005va,Berglund:2005vb}.

\section{Bubbling solutions}
\label{Sec: bubble solution}
In this section we briefly review the smooth horizonless bubbling solutions \cite{Bena:2005va,Berglund:2005vb} constructed in M-theory on a CY three-fold, which we choose to be the (non-compact) $T^4/\ZZ_2 \times \IR^2$ spanned by $(x^{5,6,7,8},x^9, x^{10})$.

All supersymmetric solutions \cite{Gauntlett:2002nw,Gutowski:2004yv,Bena:2004de,Giusto:2012gt} are a time fibration over a hyper-K\"ahler base space, which we take to be K3, and we  later approximate by the GH metric above. The four-form fluxes we  turn on lie in $H^{1,1}_-(T^4/{\mathbb Z_2})\times H^{1,1}_+(\mathrm{K3})$, where the plus (minus) sign denotes self-dual (anti-self dual) cohomology classes. With this sign convention, which follows the microstate geometry literature \cite{Bena:2004de,Bena:2015bea}, the fluxes are anti-self-dual on the 
$T^4/{\mathbb Z_2} \times$K3 compactification four-fold. This is the opposite from the convention of \cite{Becker:1996gj}, more commonly used in the flux-compactification literature.

We introduce complex coordinates and $(1,1)$-forms on a rectangular  $T^4/{\mathbb Z}_2$ with lattice identification $x^m\sim x^m+1$, and ${\mathbb Z}_2$ involution $x^m \sim -x^m$:
\begin{alignat}{2}
\label{eq:def-complex-structure-11-forms}
w_1&\!=\!L_5x^5+iL_6x^6,
&\quad
w_2&\!=\!L_7x^7+iL_8x^8, \\[-1mm]
J_1&\!\equiv\!\tfrac{i}{2}dw_1\wedge d\bar w_1,
&
J_4&\!\equiv\!\tfrac12\!\left(dw_1\wedge d\bar w_2
+d\bar w_1\wedge dw_2\right), \notag\\[-1mm]
J_2&\!\equiv\!\tfrac{i}{2}dw_2\wedge d\bar w_2,
&
J_5&\!\equiv\!\tfrac{i}{2}\!\left(dw_1\wedge d\bar w_2
-d\bar w_1\wedge dw_2\right). \notag
\end{alignat}
The form  $J_1+J_2$  is self-dual, while $J_1-J_2,~ J_4,$ and $J_5$  are anti-self-dual. We also define the two-form on the non-compact coordinates, $J_3 \equiv dx^9 \wedge dx^{10}$. The eleven-dimensional solutions descend to solutions of a five-dimensional $U(1)^5$ ungauged supergravity with intersection form
\begin{equation} 
C_{3JK}  ~=~ {\small
\begin{pmatrix} 
0&1&0&0&0\\  1&0&0&0&0\\0&0&0&0&0\\  0&0&0&-2&0\\ 0&0&0&0&-2
\end{pmatrix}   \,}.
\label{hatCform}
\end{equation}
Following \cite{Gauntlett:2002nw, Gutowski:2004yv,Bena:2004de}, the three-form gauge fields of all supersymmetric solutions of this theory are:
\begin{equation}
    C^{(3)}=A^{I}\wedge J_{I}\,,~~A^I = \tfrac12 \frac{ C^{IJK}  Z_J  Z_K}{Z^3}  \, (dt +  k) + B^I \,,
\label{eq: def one forms}
\end{equation}
where  $Z  ~\equiv~  \Big(\frac{1}{6} \,C^{IJK} Z_I\, Z_J\, Z_K  \Big)^{\frac{1}{3}}$.  The magnetic four-form field strength is 
\begin{equation}
	G_4=\Theta^I\wedge J_I \,, \quad \quad {\rm with} \  \ \Theta^I \equiv d B^I\,.
\label{G4generic}
\end{equation}
Since the solutions have nontrivial cycles wrapped by fluxes, the $B^I$ are not globally defined, but the $\Theta^I$ are.
Supersymmetry requires the $\Theta^I$ to be self-dual in the hyper-K\"ahler base, ${\bf B}$, which for us is K3, and that the $Z_I$ warp factors and the {``rotation one-form"}, ${ k} $, satisfy 
\begin{alignat}{2}
    &   d \star_4 d Z_I =  \frac{1}{2}  C_{IJK}  \Theta^J \wedge \Theta^K,  \label{eq:Z-box}  \\
   &d {k}+\star_4 d {k}=Z_I\Theta^I\, . \label{eq:k-box}
\end{alignat}
The full 11-dimensional metric with $Z_4=Z_5=0$ is\footnote{ When $Z_4$ and $ \Theta^4$ are turned on, the eleven-dimensional solutions are more complicated and can be found in \cite{Giusto:2012gt}.}:
\begin{alignat}{2}
    ds_{11}^2 &=Z_3^{-\frac{2}{3}} \!\left(-{(Z_1Z_2)^{-\frac{2}{3}}}\!\left(dt\!+\!{k}\right)^2\!+\! Z_1^{\frac{1}{3}} Z_2^{\frac{1}{3}}\! \left(dx_9^2\!+\!dx_{10}^2\right)\!\right)\nonumber\\
    &+\left(Z_1Z_3Z_2^{-2}\right)^{\frac{1}{3}}\left((L_7dx_7)^2+(L_8dx_8)^2\right)\nonumber\\
    &+\left(Z_2 Z_3 Z_1^{-2}\right)^{\frac{1}{3}}\left((L_5dx_5)^2+(L_6dx_6)^2\right)\nonumber\\
    &+\left(Z_1Z_2Z_3\right)^{\frac{1}{3}}ds_{4\,\mathbf{B}}^2\,.
    \label{eq: bubble solution metric}
\end{alignat}

It is easy to see that when 
\begin{equation}
Z_1 = Z_2=1~,~~  \Theta^3=0~,~~~  \Theta^1=- \Theta^2 \,,
\label{eq:limit}
\end{equation}
the bubbling solution \eqref{eq: bubble solution metric} describes a Becker-Becker \cite{Becker:1996gj} warped flux compactification on $T^4/{\mathbb Z}_2 \times $K3, with warp factor $Z_3$. There are two mechanisms that ensure that this bubbling solution is Lorentz invariant along $t, x_9, x_{10}$: the first is the fact that $Z_1$ and $Z_2$ are constant, and the second is that $Z_1 \Theta^1+ Z_2 \Theta^2 =0$ which, together with the vanishing of $ \Theta^3$, ensures that there is no source for ${k}$ in \eqref{eq:Z-box} and therefore $k$ can be set to zero. 

The magnetic four-form field strength \eqref{G4generic} reduces to  
\begin{equation}
    G_{4,\text{Magn}}=L_5 L_6\Theta^1 \wedge dx^5\wedge dx^6+L_7 L_8\Theta^2 \wedge dx^7\wedge dx^8\,,
\end{equation}
and is anti-self dual, since $\Theta^1=-\Theta^2$.

\section{Fixing moduli of $T^4/\ZZ_2\times \text{K3}$}

We now show more explicitly how turning on fluxes stabilizes some of the moduli of a $T^4/\mathbb{Z}_2 \times $K3 compactification, paving the way to the general solution where moduli are not fixed at their stabilized value, and Lorentz invariance is broken. Although the argument can be made very generally, we focus on a rectangular $T^4=(S^1)^4$, whose lengths are $L_5,L_6,L_7,L_8$, that we quotient by $\mathbb{Z}_2$. These lengths can be combined into dimensionless ratios, corresponding to moduli of the compactification that can, in principle, be stabilized by fluxes, and an overall volume modulus. 

\subsection*{ One modulus:}

To fix one of the dimensionless moduli  we begin with a simplified solution where $Z_4=Z_5=0$ and $\Theta^4=\Theta^5=0$. We make the argument explicit by using the GH approximation of K3 explained in Section \ref{GH-K3}, but  the same argument can be made without using this approximation.  As explained in Appendix \ref{App:K3bubbling}, the full smooth GH bubbling geometries can be written explicitly by specifying eight harmonic functions, whose poles are required to satisfy certain regularity conditions \cite{Bena:2005va,Berglund:2005vb}. 

The $U(1)$ fiber shrinks at the position of each D6 brane/GH center, therefore any path between two GH centers, $\vec{y}_i$ and $\vec{y}_j$,  generates a nontrivial two-cycle, $\Delta_{ij}$. Let $K^I$ have residues $k_i^I$ at the GH centers. The flux of $\Theta^I$ through $\Delta_{ij}$ is proportional to $k_i^I/q_i-k_j^I/q_j$. Thus, a nonzero flux through $\Delta_{ij}$ requires this difference to be nonzero.

By integrating $G_4$ along any such two-cycle $\Delta_{ij}$, and $x^5,x^6$ with unit period, we obtain the quantization condition:
\begin{equation}
    \int_{\Delta_{ij}\times T_{56}} G_4\equiv n_{56} \in \mathbb{Z}\,,\label{eq:quantization}
\end{equation}
where we dropped from $n_{56}$ the indices corresponding to the cycle $\Delta_{ij}$ to improve readability (in Appendix \ref{App:tadpole} these indices are explicitly written, see \eqref{genquantization}).
In the GH approximation this integral can be calculated exactly. For a cycle between two D6 branes with charges $q_i=q_j=1$ (see \eqref{harmonics}), we get
\begin{equation}
    \int_{\Delta_{ij}\times T_{56}} G_4=4\pi(k^1_i - k^1_j)L_5 L_6\,,\label{eq:quantization2}
\end{equation}
where $k^1_i$ are the poles of the $K_1$ harmonic function.

A similar computation for $G_4$ along $T_{78}$ leads to $4\pi(k^i - k^j)L_7 L_8 \equiv n_{78}\in \mathbb{Z}$ \footnote{There is one additional quantization condition for each flux component, coming from a K3 two-cycle that is not represented by a simple path between two GH centers, although its Poincaré dual can be written explicitly (see \eqref{swcycle}). For the two flux components considered here, and for $q_i=1$, these conditions are $\frac{1}{2}\sum_{i=1}^{16}4\pi L_5L_6 k_i^1\in\mathbb{Z}$ and $\frac{1}{2}\sum_{i=1}^{16}4\pi L_7L_8 k_i^2\in\mathbb Z$. The crucial factor of $1/2$ makes these conditions nontrivial. See Appendix \ref{App:tadpole} for more details.}. We recall that for a single D6-brane at the i\emph{th} site, $q_i=1$. The requirement of Lorentz invariance \eqref{eq:limit} implies that $K^1=-K^2$, which gives a relation between quantized fluxes and moduli:
\begin{equation}
    \frac{n_{78}}{L_7L_8}=-\frac{n_{56}}{L_5L_6}\,.
    \label{eq:stabilization}
\end{equation}
Hence, turning on these fluxes on $T^4/\mathbb{Z}_2 \times $K3 stabilizes  the dimensionless modulus $\frac{L_5L_6}{L_7L_{8_{\,}}}$. Note that this stabilization does not rely on the GH approximation of the K3 metric. The self-duality of the $\Theta^I$ inside K3 and the anti-self-duality of the fluxes in $T^4/\ZZ_2 \times$ K3 are enough to fix this modulus.

\subsection*{Fixing more moduli}

Fixing more dimensionless moduli on $T^4$  requires adding the magnetic fluxes $\Theta^4$ and $\Theta^5$. Let us first discuss a solution with $\Theta^4$ only, whose corresponding magnetic four-form is:
\begin{equation} \begin{aligned}
  \label{eq: G4 written as antiself dual}
    G_{4}&=\Theta^4 \wedge \frac{dw_1\wedge d\bar w_2 +  d\bar
  w_1 \wedge d w_2 }{2}\\
  &= \Theta^4 \wedge \left( L_6L_8 ~dx_6 \wedge dx_8+ L_5L_7 ~dx_5 \wedge dx_7 \right)  \nonumber \,,
\end{aligned} \end{equation}
which is anti-self-dual by construction since $\Theta^4$ is self-dual on K3 and the $T^4$ part is anti-self-dual. Flux quantization requires 
\begin{equation} \begin{aligned}
    \int_{\Delta_{ij}\times T_{68}} G_{4}&=-4\pi (k_i^4 - k_j^4)L_6L_8=n_{68}\in \mathbb{Z}\,,\nonumber\\
    \int_{\Delta_{ij}\times T_{57}} G_{4}&=-4\pi (k_i^4 - k_j^4)L_5L_7=n_{57}\in \mathbb{Z}\,,\nonumber \,
\end{aligned} \end{equation}
which leads to  
\begin{equation}
    \frac{n_{68}}{L_{6}L_{8}}=\frac{n_{57}}{L_{7}L_{5}}\,.
\label{eq:Theta4stab}
\end{equation} 
We can finally turn on a flux along $J_5$; it is straightforward to reproduce the same computation as for $J_4$ and obtain
\begin{equation}
    \frac{n_{58}}{L_{5}L_{8}}=-\frac{n_{67}}{L_{6}L_{7}}\,.
\end{equation}
Thus, by tuning on three types of fluxes, we fixed three dimensionless moduli of $T^4$.

It is not hard to generalize our calculation to include complex-structure deformations that correspond to twisting the $T^4$. As explained in \cite{Bena:2012ub}, these deformations give rise to other vector multiplets in five-dimensional supergravity, and the requirement that the solution be Lorentz invariant in the three non-compact directions, combined with the flux quantization in eleven dimensions, fixes these complex-structure moduli as well.

\subsection*{The tadpole sourced by the fluxes}

It was conjectured in \cite{Bena:2020xrh,Bena:2021wyr}
that the minimal value of the tadpole charge sourced by the fluxes that stabilize a given number of moduli  grows linearly with this number. We will see that this is exactly what happens here.

The four-form fluxes source an M2-brane charge 
\begin{equation}
    Q_\text{Tadpole}=-\frac{1}{2}\int_{ T^4/\mathbb{Z}_2 \times\text{K3}} G_4\wedge G_4\,,
\end{equation}
that enters in the tadpole cancellation condition. In Appendix \ref{App:tadpole} we compute this charge for our solutions. When $k_i^1=-k_i^2$ this expression becomes particularly simple:
\begin{equation}
\begin{aligned}
\label{eq:tadpole}
	Q_{\text{Tadpole}} =& ~\frac{(4\pi)^2{\text{Vol}}(T^4)}{2}\, \sum_{I=1,4,5}\sum_{i=1}^{16}\left(k_i^I \right)^2 \\
	=& ~\frac{1}{2}\sum_{I=1,4,5}\sum_{i=1}^{16} (n_i^I)^2\,,
\end{aligned}
\end{equation}
where $n_i^I=4\pi k_i^I (L_mL_n)$, are the quantized versions of the $k_i^I$, and either satisfy  $n_i^I \in \ZZ  +\frac{1}{2}$ for all $n_i^I$ or satisfy  $n_i^I\in \ZZ  $ for all $n_i^I$. More precisely, they are in the weight lattice of $\text{Spin}(32)/\ZZ_2$. 
The factor of $\frac{1}{2}$ in the quantization condition distinguishes two different configuration of fluxes, which can be traced back to the relative parity of the fluxes along the last two cycles. Indeed, if $n_{15,16}^I=n_{15,16'}^I ~(\text{mod}~ 2)$, then all the $n_{i}^I\in \ZZ$. In contrast, if $n_{15,16}^I=1+n_{15,16'}^I ~(\text{mod}~ 2)$ then the quantization condition is $n_{i}^I\in \ZZ+\frac{1}{2}$. 
See Appendix {\ref{App:tadpole} for a longer discussion. 

This equation has several important physics consequences. The first is that the quantization conditions imply that the shortest allowed nonzero vector has $\sum_i (n_i^I)^2=2$. Substituting such a vector in \eqref{eq:tadpole} gives a tadpole of $1$. The smallest tadpole is obtained by choosing the $n_i^I$ to be  $\pm 1$ at two centers\footnote{Choosing $n_i^I=\pm 1$ at only one center is incompatible with the flux quantization on the chiral spinor weight class \eqref{quantization2}.}. For each flux component $\Theta^I$ ($I=1,4,5$) that is turned on, one fixes one particular dimensionless modulus corresponding to a $T^4/\ZZ_2$ deformation. We thus have
\begin{equation}
  Q_{\text{Tadpole}}^{\rm min}= \, n_{\rm moduli\, fixed} \ .
\end{equation} 

As we explained above, it is not hard to generalize this argument to other moduli corresponding to twisting the $T^4/\ZZ_2$, and find the same linear growth. 

There is also a negative contribution to this tadpole, given by $Q_{\rm Tapole}=-\frac{\chi(T^4/\mathbb Z_2 \times \text{K3})}{24}=-24$. This charge can also be obtained directly by computing the induced D2-charge of the D6 branes and O6 planes whose uplift gives K3, which is proportional to the Euler characteristic of the cycle they wrap, $T^4/\ZZ_2$. In the fundamental domain, each of the 16 D6 branes acquires a negative D2 charge equal to $-1$, and each of the 8 O6 planes acquires a negative D2 charge equal to $-1$ \cite{Dasgupta:1997cd} so the total contribution to the D2 tadpole charge is -24.

To cancel the tadpole, one can either add extra M2 branes, or add extra fluxes that stabilize more of the $T^4/\ZZ_2$ moduli, or fine-tune the quantized fluxes so that the sum in \eqref{eq:tadpole} is equal to $24$.  As an example of fine-tuning, taking $n_i^I=(2,2,2,2,0,\dots,0)$ for each $I=1,4,5$ gives $\sum_i(n_i^I)^2=16$ for each of the three flux components, and hence cancels the negative tadpole coming from the Euler characteristic. More generally, any allowed choice of flux vectors satisfying $\sum_{I=1,4,5}\sum_i(n_i^I)^2=48$ gives the same total tadpole.

The $T^4\times$K3 version of our construction can be straightforwardly dualized to a Type IIB String Theory flux compactification on $T^2 \times $K3, by first reducing to Type IIA along one of the $T^4$ directions (say $x^5$) and T-dualizing along another one (say $x^7$). The structure of the warp factors in \eqref{eq: bubble solution metric} ensures that, in the absence of $\Theta^{5}$, the IIB compactification is a warped product and has $3+1$-dimensional Lorentz invariance. The fluxes proportional to $\Theta^1, \Theta^2$ and $\Theta^4$ become ISD complex-three-form fluxes. Furthermore, the equation for the D3 tadpole is the same as the equation for the M2 tadpole in M-theory \eqref{eq:tadpole}, giving the same linear relation between the number of stabilized moduli and the minimum D3 tadpole.

\section{Relaxing Lorentz invariance and moduli destabilization} 

In the previous section we imposed 2+1 dimensional Lorentz invariance along the non-compact directions, and we have seen that this is obtained by demanding that the four-form field strength be (anti) self-dual in the eight-dimensional compact space. However, we have also seen that there are many solutions that do not preserve 2+1  Lorentz invariance and a warped-product structure, and which have fluxes that extend along some, but not all of the space-time directions. 

It is not hard to construct such solutions using the technology of \cite{Bena:2005va,Berglund:2005vb}. As reviewed in  Appendix \ref{App:K3bubbling}, the solutions are determined by 8 harmonic functions. The first harmonic function, $V$, encodes the distribution of GH centers, \eqref{eq: metric uplift D6 brane}, the $K^{I}$ harmonic functions encode the fluxes on the two-cycles running between these centers and, for smooth horizonless solutions, the poles of the other harmonic functions are entirely determined by the poles of $V$ and those of $K^{I}$ \cite{Bena:2005va,Berglund:2005vb}. Furthermore, there is another condition that the locations of the centers must satisfy, given in \eqref{eq:bubble equation}, that comes from requiring no Dirac-Misner strings. When compactified to ten dimensions along the GH fiber, these solutions become a certain subset of the multi-center solutions of \cite{Bates:2003vx}. 

The simplest Lorentz-breaking solution continuously connected to the Lorentz-invariant one has $\Theta^4=\Theta^5=0$, and involves two D6-image pairs. We denote by $H_i$ and $H_{i'}$ the harmonic functions with unit poles at the D6 center, $\vec y_i$, and its orientifold image, $\vec y_{i'}$, and define
\begin{equation}
	D_i \equiv H_i-H_{i'} \,, \qquad S_i \equiv H_i+H_{i'} .
\end{equation}
The nontrivial harmonic functions are
\begin{equation}
	K^1=k_1^1D_1\,,\qquad K^2=k_1^2 D_1\,,\qquad K^3=k_2^3 D_2 \,,
\end{equation}
together with
\begin{equation}
L_1=L_2=1 \,, \qquad L_3=1-k_1^1 k_1^2 S_1\,,\qquad M=0 \,.
\end{equation}
The function $V$ is the same as in \eqref{eq: harmonic function K3}, since it encodes the K3 geometry rather than the fluxes. The functions $K^I,L_I$ and $M$ have no poles at the O6 centers and, for simplicity, we focus on solutions where they have no poles at the remaining GH centers.  

Using \eqref{Zkform} and \eqref{eq:angularmomentum}, the warp factors and the angular-momentum function, $\mu$, are
\begin{align}
	Z_1&=Z_2=1+\frac{k_1^2k_2^3\,D_1D_2}{V}\,, 
	\nonumber \\
	Z_3&=1-k_1^1k_1^2\,S_1+\frac{k_1^1k_1^2\,D_1^2}{V} \,, \label{runningZ} \\
	\mu&=\frac{(k_1^1+k_1^2)\,D_1+k_2^3\,D_2(1-k_1^1k_1^2\,S_1)}{2V} +\frac{k_1^1k_1^2k_2^3\,D_1^2D_2}{V^2} \,.  \nonumber
\end{align}

The bubble equations \eqref{eq:bubble equation} of the two GH-image pairs reduce, by the orientifold symmetry, to the  following two equations:
\begin{align}
	k_1^1k_1^2k_2^3\left(H(\vec{y}_1-\vec{y}_2)-H(\vec{y}_1-\vec{y}_{2'}) \right)&=-(k_1^1+k_1^2)\,, \label{be1} \\
	k_1^1k_1^2k_2^3\left(H(\vec{y}_1-\vec{y}_2)+H(\vec{y}_1-\vec{y}_{2'}) \right)&=k_2^3\,. \label{be2} 
\end{align}
We use \eqref{be1} to determine $k_2^3$:
\begin{equation}
	k_2^3=-\frac{k_+}{k_1^1k_1^2 \left(H(\vec{y}_1-\vec{y}_2)-H(\vec{y}_1-\vec{y}_{2'}) \right)} \,,
\end{equation}
where we introduced the parameter 
\begin{equation}
	k_+\equiv k_1^1+k_1^2 \,.
\end{equation}
This parameter measures the departure from the Lorentz-invariant solution: when $k_+=0$, one has $K^1=-K^2$ and \eqref{be2} forces $k_2^3=0$, hence $K^3=0$, so that the solution reduces continuously to the Lorentz-invariant compactification of the previous section (note that $k_2^3$ is not subject to any quantization condition). Away from this limit, the source term $Z_I \Theta^I$ in the equation for the rotation one-form \eqref{eq:k-box} becomes nontrivial: the solution acquires a finite rotation vector, $k$, and the metric has a component mixing the time direction with the compact space. The full eleven-dimensional solution can therefore no longer be written as a warped-product compactification. 

The same parameter, $k_+$, also describes the departure from the (anti) self-duality of the fluxes:
\begin{equation}
	\star_8 G_4 + G_4 = k_+ \partial_a \left(\frac{D_1}{V} \right)\Omega^{a,+}\wedge(J_1+J_2) \, ,
\end{equation}
where $\Omega^{a,+}$ are the self-dual two forms on $T^4$ introduced in \eqref{eq:6cycles}. Note that the parameter $k_+$ is physically meaningful: it controls the deviation of $G_4$
from anti-self-duality, which is a gauge-invariant statement, and hence the Lorentz-breaking deformation is not a gauge artifact. This is similar to what happens in \cite{Butti:2004pk}. The difference is that in those solutions Lorentz invariance is preserved, but the Killing spinors are deformed, while in our solutions Lorenz invariance is broken but the Killing spinors are the same as in the solution with anti-self-dual fluxes.

Using the quantization conditions \eqref{quantki}, the parameters $k_1^1$ and $k_1^2$  are related to quantized fluxes by
\begin{equation}
	k_1^1=\frac{n^1_{1,56}}{4\pi L_5L_6}\,, \qquad k_1^2=\frac{n^2_{1,78}}{4\pi L_7L_8} \,,
\end{equation}
and it is evident that $k_+=0$ gives \eqref{eq:stabilization}. However, for $k_+\neq0$, this relation is no longer true. Instead, on the Lorentz-breaking branch the second bubble equation \eqref{be2} gives 
\begin{equation}
	n_{56}n_{78} \left(\!H(\vec{y}_1\!-\!\vec{y}_2)+H(\vec{y}_1\!-\!\vec{y}_{2'}) \right)=(4\pi)^2 {\rm vol}(T^4) \,, \label{stabilized}
\end{equation}
where we used the simplified notation of \eqref{eq:stabilization} for the quantized fluxes\footnote{Note that if one deforms the Lorentz-invariant solutions keeping the same quantized fluxes, $n_{56} n_{78} <0$, and the equation above has solutions because the function $H$ is negative in certain regions \eqref{eq: harmonic function T3}. However, one can also try to construct directly Lorentz-breaking solutions with $n_{56} n_{78} >0$}.

Therefore, after relaxing the warped-product ansatz, the stabilization of the $T^4/\ZZ_2$ dimensionless modulus $\frac{L_5L_6}{L_7L_8}$ is replaced by a constraint involving quantized fluxes, the torus volume and the positions of the D6 branes that enter as K3 moduli. 

Hence, at the Lorentz-invariant point, the $\Theta_1$ and $\Theta_2$ fluxes stabilize one of the dimensionless $T^4/\ZZ_2$ shape moduli, and leave massless the 48 moduli corresponding to D6 positions inside $T^3/\ZZ_2$. Away from the Lorentz-invariant point, these fluxes do not stabilize this $T^4/\ZZ_2$ shape modulus anymore, but constrain the $T^4/\ZZ_2$ volume modulus and one combination of the 48 D6 position moduli to satisfy \eqref{stabilized}, thus leaving 47 massless position moduli. The fluxes are no longer anti-self-dual, and  the nature of the stabilized modulus changes:  it is now a combination 
involving the volume modulus, which cannot be stabilized by fluxes in the EFT, making this a genuinely eleven-dimensional effect invisible to the 
effective description.

 Note that total number of moduli stabilized by the same fluxes can change between the Lorentz-breaking and the Lorentz-invariant compactifications. Adding only $\Theta_1=-\Theta_2$ on more cycles of K3 will stabilize only $L_5 L_6/L_7 L_8$ in the Lorentz-invariant compactification, but will give rise to more bubble equations similar to \eqref{stabilized}, and will therefore stabilize more combinations of the $T^4/\ZZ_2$ volume modulus and the D6 position moduli. Remarkably, the tadpole sourced by these fluxes also grows linearly with the number of stabilized moduli. Hence, the Tadpole Conjecture holds in both the Lorentz-invariant and the Lorentz-breaking compactifications.

\section{Discussion}

We have  constructed a fully back-reacted M-theory compactification on $T^4/\ZZ_2 \times $K3 with four-form fluxes, and shown explicitly how fluxes stabilize some of the $T^4/\ZZ_2$ shape moduli in a warped-product compactification that preserves 2+1 dimensional Lorentz invariance. From the perspective of the 2+1 dimensional effective theory, which is obtained assuming a (warped) product ansatz, this is the celebrated stabilization of moduli by fluxes, which one can derive from the Gukov-Vafa-Witten superpotential \cite{Gukov:1999ya}.

However, a surprise awaits. We have also found that the solution with stabilized moduli  belongs to a larger family of perfectly regular 11-dimensional supersymmetric solutions that generically break the warped-product structure and the 2+1 dimensional Lorentz invariance, and whose $T^4/\ZZ_2$ shape moduli are not stabilized. Furthermore, these solutions do not have a self-dual four-form flux. There is nothing special that happens when these moduli reach the ``stabilized'' values, except that the solution becomes a warped product and 2+1 dimensional Lorentz invariance is restored.

Our result indicates that, at least for $T^4/\mathbb{Z}_2 \times \text{K3}$ compactifications, the EFT description of moduli stabilization by fluxes is an artifact of assuming a warped-product structure and of working in a low-energy effective theory blind to deformations away from this ansatz. If this extends to other compactifications, it would call into question one of the cornerstones of string phenomenology: the EFT description of flux-induced moduli stabilization does not reflect the dynamics of the full eleven-dimensional theory.

That said, our result does not imply that fluxes fail to stabilize any moduli. The full eleven-dimensional solutions show that fluxes do stabilize moduli, but not the ones identified by the Lorentz-invariant warped-product description. Instead, the stabilized moduli are  nontrivial combination of K3 shape moduli and the $T^4/\mathbb{Z}_2$ volume modulus, which is precisely the modulus that flux stabilization in the EFT cannot reach. Fluxes do their job, but on a modulus that the effective description is structurally blind to.

To prove that our result is universal several steps are needed. The first is to ascertain whether the supersymmetric Lorentz-breaking solutions that we obtained using the Gibbons-Hawking approximation of the K3 metric are valid on the exact K3 metric. Indeed, the GH approximation only captures the K3 metric away from the AH centers corresponding to (the uplift of) O6-planes. Furthermore, the GH $U(1)$ isometry direction, which mixes with the time direction in the Lorentz-breaking solutions, is, in fact, not an exact isometry of the AH metric or of the full K3 geometry. However, the fact that a certain direction is not an isometry direction does not prevent the existence of supersymmetric solutions where it mixes with time. There are numerous examples where this happens \cite{Bena:2007kg}.

In this Letter we have checked explicitly that the Lorentz-breaking solutions have the same regularity properties as the Lorentz-invariant ones away from the AH centers, where the GH approximation is valid. Furthermore, we expect the Lorentz-invariant solutions to extend trivially to the region of K3 close to the AH centers.

Thus, it is crucial to demonstrate that Lorentz-breaking solutions are also valid in the vicinity of the AH centers. There are two ways of doing this. One is more formal, and consists in showing that, for the Lorentz-breaking solution, the functions $Z_1$ and $Z_2$ remain smooth (as one can see intuitively from \eqref{runningZ}), and that, moreover, the rotation vector ${k}$ that they source does not develop pathologies in this region. The second is to solve explicitly the equations (\ref{eq:Z-box}) in an AH space, as was done in \cite{Bena:2007ju}, and to match these solutions in the region where the AH metric and the GH metric overlap via a matched asymptotic expansion. 

Assuming that the Lorentz-breaking solution remains regular near the AH centers, one can ask whether our result is generic, or just an artifact of using a simple compactification manifold. To answer this we can first try to extend our work to K3$ \times $K3 compactifications, and then to more general CY 4-folds.  Another route for ascertaining  whether our result is generic is to dualize our construction to a IIB K3$\times T^2$ compactification (or an orbifold thereof) and possibly extend our calculation to more involved CY 3-folds. 

Our result explicitly confirms the Tadpole Conjecture \cite{Bena:2020xrh} both for Lorentz-invariant  and for Lorentz-breaking compactifications. Although the nature of the stabilized moduli is different, the minimal tadpole of the fluxes that need to be turned ongrows linearly with the number of stabilized moduli in both cases.
 It would be very interesting to relate our construction with the Lorentz-invariant K3$\times $K3 analysis of \cite{Bena:2020xrh,Bena:2021wyr}. It would also be interesting to extend our analysis to more general deformations of $T^4/\ZZ_2$. The five-dimensional supergravity solutions corresponding to these compactifications would have the same K3 base, but have extra  vector multiplets \cite{Bena:2012ub}, corresponding to $T^4/\ZZ_2$ complex-structure moduli. We expect that turning on these fluxes and imposing a warped-product Lorentz-invariant structure will ``stabilize'' all moduli but the volume, at the cost of a contribution to the tadpole sourced by these fluxes equal to $1$ per extra stabilized modulus.  It would also be interesting to construct the Lorentz-breaking solutions with fluxes that wrap more complicated cycles of K3. These fluxes will give rise to more equations of type \eqref{stabilized} which will stabilize more combinations of D6-brane position moduli with the $T^4/\ZZ_2$ volume modulus. An important question is how many such combinations can be stabilized with the fluxes visible in the GH approximation of K3.

The structure of the Lorentz-breaking supersymmetric solutions, the role of $J_1+J_2$ in controlling the deviation from anti-self-duality, and the mixing of the time direction with the internal space, are reminiscent of the $\Omega$ background of Nekrasov \cite{Nekrasov:2002qd}. It would be interesting to make this connection more precise. 

 Last but not least, perhaps the most important question raised by our construction, is whether our Lorentz-breaking solutions  can be captured in any low-energy effective theory. For example, these solutions have a nontrivial profile on the compactification manifold for a metric component with $g_{ti}$, that has one leg along time and one leg along the compactification manifold. Since the compactification manifold does not have nontrivial one-cycles, this field should not be  described by a low-energy EFT. However, it may be possible that integrating out this field will result in a new metric in which $g_{tt}$ is renormalized by a multiplicative constant, and still give a Lorentz-invariant background that has an EFT description. It would be fascinating to see whether this can happen.

\noindent {\bf Acknowledgments:} We would like to thank Per Berglund, Antoine Bourget, Ang\`ele Lochet, Severin L\"ust, Emil Martinec, Ruben Minasian, Miguel Montero, Miguel Morros, Radu Roiban, Ashoke Sen, Cumrun Vafa, Johannes Walcher and Nick Warner for valuable discussions. The work of DT is supported by the Israel Science Foundation (grant No. 1417/21), by the German Research Foundation through a German-Israeli Project Cooperation (DIP) grant ``Holography and the Swampland,'' by Carole and Marcus Weinstein through the BGU Presidential Faculty Recruitment Fund, by the ISF Center of Excellence for theoretical high energy physics, by the VATAT Research Hub in the Field of quantum computing, by a scholarship by the Ministry of Foreign Affairs of Israel and by the ERC starting Grant dSHologQI (project number 101117338).

\begin{widetext}
\appendix 
\section{Bubbling solutions on K3 surfaces}
\label{App:K3bubbling}

\subsection*{K3 in the Gibbons-Hawking approximation}

The Kummer construction produces a K3 surface by taking the quotient $T^4/\mathbb Z_2$, where the $\mathbb Z_2$ acts by simultaneous sign flip on all four coordinates, and blowing up the resulting 16 orbifold singularities. Each blow-up contributes an exceptional divisor, $\mathbb P^1$. These 16 two-cycles, together with the 6 two-cycles inherited from the torus, span a sublattice of $H_2(K3,\mathbb Z)$. The extra cycles needed to complete this to the full, unimodular lattice $H_2(K3,\mathbb Z)\cong \Gamma^{3,19}$,
are half-integer linear combinations of the 16 cycles above, and their existence is required by the self-duality of the lattice.

A natural way to see a rank-16 part of the K3 lattice is through the M-theory uplift of type IIA on $T^3/\mathbb Z_2$ orientifold with 16 D6 branes and 8 O6 planes \cite{Sen:1997kz,Schulz:2012wu}. The $\mathbb Z_2$ acts by simultaneous sign flip on the three torus coordinates and the orientation, and has 8 fixed points, corresponding to the O6 planes, as shown in Figure \ref{fig:K3-GH}. The M-theory lift of a D6 brane is a Taub-NUT (GH) center, while the lift of an O6 plane is an Atiyah-Hitchin space. Far from the O6 planes, the Atiyah-Hitchin metric is well approximated by a negative-charge Taub-NUT (GH) center. In this approximation the K3 metric takes the Gibbons-Hawking form \eqref{eq: metric uplift D6 brane} with $T^3/\mathbb Z_2$ base and with $V$ given by \eqref{eq: harmonic function K3}. Since the orientifold action flips the orientation of the  M-theory direction, $\psi$, the GH $U(1)$ is only an approximate isometry of the region where the Gibbons-Hawking approximation is valid, and is broken near the O6 (AH) centers. Hence, it is not an exact isometry of the full K3 metric.

The GH fiber shrinks at every D6 center and hence a path in the three-dimensional base between two GH centers gives a two-sphere in the four-dimensional space. Choosing an ordering of the 16 D6 centers, we take the 16 independent cycles to be the 15 cycles, $\Delta_{i,i+1}$, running from center $\vec y_{\,i}$  to center $\vec y_{\,i+1}$, together with one additional cycle ${\Delta}_{15\overline{16}}$ running from the 15th center to the image of the 16th center. We label these cycles by
\begin{equation}
        \Delta_A\equiv 
        \begin{cases} 
        \Delta_{i,i+1}\,, \  A=1,\dots ,15 \  ({\rm and} \, i=1,\dots ,15) \, , \,  \\
        {\Delta}_{15\overline{16}} \ , \ A=16 \,.
	\end{cases}
	\label{2cycles}
\end{equation}
These cycles generate the $D_{16}$ root sublattice of $H_2(K3,\mathbb Z)$, spanning all two-cycles obtained from paths connecting two GH centers, corresponding to two D6 branes, two image D6 branes, or any cycle between a brane and an image brane. With the orientation convention used in the microstate-geometries literature, their intersection matrix is the Cartan matrix $I_{AB}$ of $SO(32)$ with a plus sign, which is opposite to the usual K3 convention. Their Poincar\'e-dual two forms are
\begin{equation}
\begin{aligned}
       \omega^A = 
       \begin{cases}
       \partial_a \left(\frac{G_A-G_A'}{V}\right)\Omega^{a,+} \,, \quad  G_A\equiv H_i - H_{i+1} \, ,  \quad A=1,\dots,15 \ ({\rm and} \, i=1,\dots ,15) \,, \\
	\partial_a \left(\frac{G_{16}-G_{16}'}{V} \right)\Omega^{a,+}\, ,  \ \  G_{16} \equiv H_{15}-H'_{16} \,, \quad \ A=16 \,,
	\end{cases}
\end{aligned}
\label{omega^A}
\end{equation} 
where $H_i$, $H_i'$ denote the harmonic functions sourced respectively by the $i$th D6 GH center and its orientifold image (see more details on these functions in the next subsection). These forms satisfy
\begin{equation}
	\int_{K3}\, \omega^A \wedge \omega^B = I^{AB} \,, 
\label{InverseIntersection}
\end{equation}
where $I^{AB}$ is the inverse of $I_{AB}$.

As already mentioned, these two-forms do not  generate by themselves the full integral $H_2(K3,\ZZ)$ cohomology lattice. Besides the forms inherited from the $T^4$ that we write below in \eqref{eq:6cycles}, the K3 lattice contains an additional (chiral) spinor weight class that completes the $D_{16} (\subset \Gamma^{3,19})$ lattice to the ${\rm Spin}(32)/\mathbb Z_2$ one. The corresponding two-forms are generated by
\begin{equation}
\label{swcycle}
    \omega_{\text{sp}}=\frac{1}{2} \sum_{i=1}^{16} \epsilon_i \, \partial_a \left(\frac{H_i-H_i'}{V}\right)\Omega^{a,+}  \,  \,, \quad \epsilon_i= \pm 1 \, , \quad \ {\rm with} \ \prod_{i=1}^{16}\epsilon_i=+1 \ .
\end{equation}

Finally, the six two-forms inherited from the $T^4$ in the Kummer $T^4/\mathbb Z_2$ construction of K3 are
\begin{equation}
        \Omega^{a,\pm} = e^1\wedge e^{a+1} \mp {1\over 2}\epsilon_{abc}\, e^{b+1}\wedge e^{c+1} \,, \qquad a=1,2,3 \,,
\label{eq:6cycles}
\end{equation}
with 
\begin{equation}
        e^1 = V^{-1/2}(d\psi+A) \,, \qquad e^{a+1}=V^{1/2}dy^a \,.
\end{equation}
In our conventions, $\Omega^{a,+}$ are self-dual, while $\Omega^{a,-}$ are anti-self-dual. The latter define the hyper-K\"ahler structure of K3.

\begin{figure}[h]
    \centering
    \includegraphics[width=0.47\linewidth]{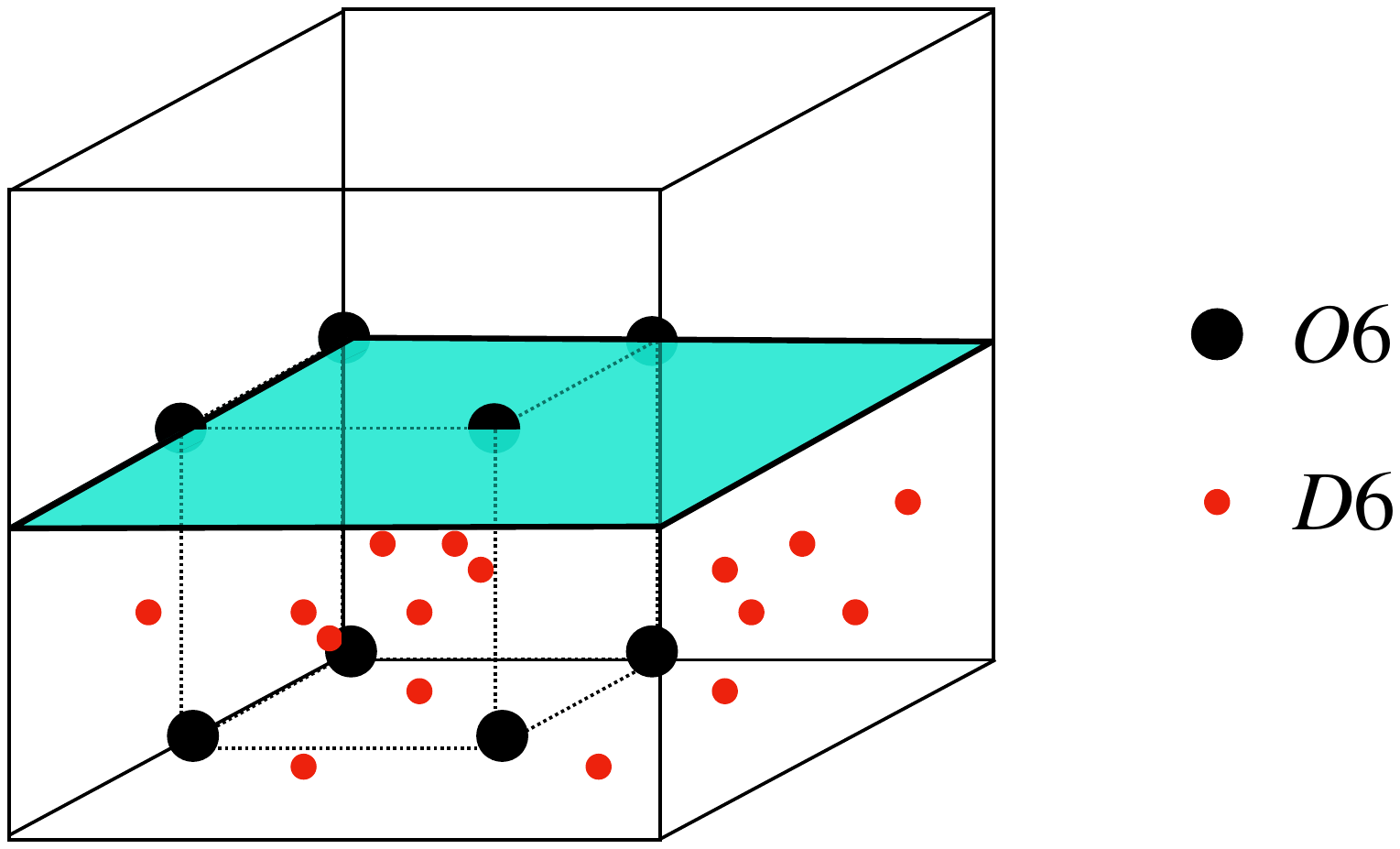}
    \caption{A representation of the $T^3/\ZZ_2$ space. The eight O6 planes (in black) sit at the fixed points of the $\ZZ_2$ involution, and the sixteen D6 branes (in red) are free to move everywhere inside the fundamental domain, which is below the blue plane. There are also 16 image D6 branes, above the blue plane, which we do not represent.}
    \label{fig:K3-GH}
\end{figure}

\subsection*{Details on bubbling solutions}
\vspace{-1mm}

When the four-dimensional base space has the Gibbons-Hawking form \eqref{eq: metric uplift D6 brane}, the bubbling solutions are completely determined in terms of a set of harmonic functions  $(V,K^I,L_I,M)$ on the flat three-dimensional (locally-$\IR^3$) base, $\mathcal{B} \subset {\bf B}$, of this Gibbons-Hawking space, with coordinates $y^a$  \cite{Gauntlett:2004wh,Bena:2005ni}. The relevant fields are

\begin{equation} \begin{aligned}
    \Theta^I~=~&\partial_a \left(\frac{K^I}{V}\right) \Omega^{a,+}\,,\quad k ~=~ \mu\, ( d\psi  + \vec A \cdot d\vec y   ) ~+~ \vec \omega \cdot \, d†\vec y \,,  \\
    Z_I ~=~ &\frac{1}{2}  ~ C_{IJK} ~ \frac{K^J K^K }{V} ~+~ L_I\,, 
\label{Zkform}
\end{aligned} \end{equation} 
where the self-dual two-forms $\Omega^{a,+}$ are defined in \eqref{eq:6cycles}, and 
\begin{equation}
\begin{aligned}
\mu ~=~ &\frac {1}{6} \, C_{IJK}\,  \frac{K^I K^J K^K}{ V^2} ~+~
\frac{ K^I L_I}{2 \,V}  ~+~  M \,, \\
\vec \nabla\times  \vec \omega ~=~ & V \vec \nabla M  ~-~
M \vec \nabla V ~+~   \frac{1}{2}\, (K^I  \vec\nabla L_I - L_I \vec
\nabla K^I )\,.
\label{eq:angularmomentum}
\end{aligned}
\end{equation}

In order to obtain a smooth horizonless solution, all the functions must be harmonic on the three-dimensional base and must be writable as sums of Green functions with poles at the GH centers. In the K3 Gibbons-Hawking approximation the physical base is $T^3/\mathbb Z_2$, but it is convenient to use the Green function on $T^3$, with the $\mathbb Z_2$ identification implemented by including the image centers. We denote by $H_i\equiv H(\vec y-\vec y_i)$ the Green function with a pole at $\vec{y}_i$. Our main interest is in a compact $T^3$ base space for which the Green function satisfies
\begin{equation}
\label{eq: harmonic function T3}
-\nabla^2H(y,y_i)=\delta(y-y_i)-\frac{1}{\text{Vol}(T^3)}\,,\qquad  \ \int_{T^3}d^3y \, H(y,y_i)=0\,.
\end{equation}
The term $-\frac{1}{\text{Vol}(T^3)}$ ensures that $-\nabla^2H(\vec{y},\vec{y}_i)$ integrates to zero, as it must on a compact space, while the second integral fixes the integration constant. We can then write
 \begin{equation}
 \begin{aligned}
 \label{harmonics} 
    V &= q_0+\sum_i q_{i}H_i\,,\qquad  K^{I} = k_0^{I}+\sum_i k_i^{I}H_i  \,,\\
    L_{I} &= l_0^I + \sum_i l_i^I H_i\,,\qquad M =m_{0}+ \sum_i m_{i} H_i\,,
\end{aligned}
\end{equation}
where  $q_i=1$ or $-4$ respectively for a D6-brane and O6 plane. 
Note that, by construction, $\sum_i q_i=\sum_i k_i^I=0$ in all the geometries considered in this paper, while $\sum_i l_i^I \neq 0$. The vanishing of the integral of $H$ in \eqref{eq: harmonic function T3} therefore gives an unambiguous definition of $l_0^I$. 
The $\ZZ_2$ action flips the orientation of M-theory circle, and therefore imposes that the $K^I$ and $M$ harmonic functions (corresponding to D4 and D0 charges) are odd under the $\ZZ_2$ flip. Hence, $k_0^{I}=m_0=0$,  $k_i^I=-k_{i'}^I$ and $m_i=-m_{i'}$. Furthermore, the coefficients $l_i^{I}$ and $m_{i}$  should satisfy certain constraints that ensure that $Z_I$ and $\mu$ have no poles at any center \cite{Bena:2005va,Berglund:2005vb}:
\begin{equation}
    l^I_{i}\,=\,-\frac{1}{2q_i}C_{IJK} k^J_i k^K_i \,,\qquad 
    m_i=\, 
    \frac{1}{12q_i^2}C_{IJK}k^I_ik^J_i k^K_i \,.
\end{equation}
In addition, the absence of Dirac-Misner strings requires  \cite{Bates:2003vx,Bena:2005va,Berglund:2005vb}
\begin{equation}
    \sum_{j \ne i} 
    \langle\Gamma_{i},\Gamma_{j}\rangle
     H(|\vec y_i - \vec y_j|)=\langle \Gamma_{0},\Gamma_{i}\rangle \,,
    \label{eq:bubble equation}
\end{equation}
where the symplectic product of the charge vectors $\Gamma_{i}\equiv \left(q_i,k^{I}_i,l^{I}_i,m_i\right) $ (and similarly for $\Gamma_0$) is 
\begin{equation}  
\langle\Gamma_i,\Gamma_{j}\rangle\, =\, q_i m_{j}- q_{j}m_i+ \frac{1}{2} \sum_I
    \left(k_i^{I} l_j^{I}-k_j^{I}l_i^{I}\right).
\end{equation}
The ``bubble equation'' \eqref{eq:bubble equation} constrains the relative positions of the poles of the harmonic functions in terms of the fluxes on the two-cycles between these poles.

\section{Tadpole computation}
\label{App:tadpole}

In this appendix, we derive the flux contribution to the M2-brane tadpole, given by \begin{equation}
	Q_{\text{Tadpole}}=-\frac12\int_{K3\times T^4/\mathbb{Z}_2}G_4\wedge G_4 \,.
\end{equation}
Note that we use the conventions of \cite{Gibbons:2013tqa}, which differ by an overall sign from the ones typically used in the flux compactification literature.

The four-form field strength is given by 
\begin{equation}
	G_4=\Theta^I\wedge J_I \ .
\end{equation}
For the Lorentz-invariant solutions, we must impose
\begin{equation}
	\Theta^1=-\Theta^2 \,.
\end{equation}
Thus the $I=1,2$ part of the flux is proportional to the anti-self-dual two-form $J_1-J_2$ on $T^4$. Since $J_4$ and $J_5$ are also anti-self-dual on $T^4$, the full magnetic flux is anti-self-dual on $T^4/\mathbb{Z}_2 \times$ K3 
\begin{equation}
	G_4=-\star_8 G_4 \, ,
\end{equation}
and the contribution to the tadpole is positive.

We will compute the tadpole using the explicit expression for the fluxes obtained in the GH approximation, but since this calculation only uses topological data, it can also be done without using these explicit representatives. We get 
\begin{equation} \begin{aligned}
	Q_{\text{Tadpole}} &=-\frac{1}{2}\int_{K3}\Theta^I\wedge\Theta^J \int_{T^4/\mathbb{Z}_2}J_I\wedge J_J =-\frac{1}{2}\,\frac{{\text{Vol}}(T^4)}{2}   \int_{K3}\Theta^I\wedge\Theta^J\,C_{3IJ}\\
	&=\frac{{\text{Vol}}(T^4)}{2} \int_{K3}\left( -\Theta^1\wedge\Theta^2 +\Theta^4\wedge\Theta^4 +\Theta^5\wedge\Theta^5 \right) \,.
\label{eq:tadpoleFirst}
\end{aligned} \end{equation}
For a GH center at location $i$ and another at location $i+1$, we get 
\begin{equation} \label{inttheta}
	\int_{\Delta_A}\, \Theta^I =  4\pi \kappa_A^I \,, 
\end{equation}
with
\begin{equation}
\begin{aligned}
	\kappa_A^I&\equiv \frac{k_i^I}{q_i}-\frac{k_{i+1}^I}{q_{i+1}}\,,\quad  {\rm for } \ \ A=1,\dots,15 \,, \\ 
	\kappa_{16}^I&\equiv \frac{k_{15}^I}{q_{15}}+\frac{k_{16}^I}{q_{16}}\, ,
	\label{kappa}
\end{aligned}
\end{equation}
where  in the last equality we have used the fact that the orientifold action requires $k_{i'}=-k_i$. 
Equivalently,
\begin{equation}
	\Theta^I=4\pi\,\kappa_A^I\omega^A\,.
\label{eq:theta-kappa}
\end{equation}
Using \eqref{InverseIntersection}, we find 
\begin{equation}
	\int_{K3}\Theta^I\wedge\Theta^J =(4\pi)^2\,I^{AB}\kappa_A^I\kappa_B^J\,.
\end{equation}

Substituting this into \eqref{eq:tadpoleFirst} gives
\begin{equation}
	Q_{\text{Tadpole}} = \frac{(4\pi)^2{\text{Vo}l}(T^4)}{2}\, \left[\sum_{I=4,5}\, \sum_{i=1}^{16}\left(k_i^I \right)^2 -\sum_{i=1}^{16} k_i^1 k_i^2\right] \,,
\end{equation}
where we are only turning on fluxes between GH centers corresponding to D6-branes, $q_i=1$. 

For Lorentz-invariant solutions, $k_i^2=-k_i^1$, and hence
\begin{equation}
	Q_{\text{Tadpole}} = \frac{(4\pi)^2{\text{Vol}}(T^4)}{2}\, \sum_{I=1,4,5}\, \sum_{i=1}^{16}\left(k_i^I \right)^2 \,.
\label{eq:tadpoleki}
\end{equation}
This is the same expression as the sum of the poles of the $L_3$ harmonic function, whose asymptotic behavior coincides in our solutions with that of $Z_3$> This happens because the anti-symmetry of the $K^I$ makes the first term in the expression of $Z_3$ \eqref{Zkform} subleading compared to the second one.

It remains to write this in terms of quantized flux numbers. Flux quantization requires that for any four-cycle in K3$\times T^4/\mathbb{Z}_2$ composed of a two-cycle  $\Delta_A$ on K3 and a two-cycle of $T^4/\mathbb{Z}_2$, the period of $G_4$ is an integer. The two-cycles inherited from the covering $T^4$ fall into two classes: a generic two-torus together with its $\mathbb{Z}_2$ image, and a two-torus that is already $\mathbb{Z}_2$-invariant. Fluxes through $\mathbb{Z}_2$-invariant two-tori may take fractional values, since branes wrapping such cycles are tied to the $\mathbb{Z}_2$-invariant subspace. This is analogous to the D5-branes of the Klebanov--Strassler solution, which are forced to sit at the tip of the throat, where the $S^2$ shrinks, and give therefore rise to fractional D3 charge \cite{Klebanov:2000hb}. Thus the quantization conditions arise only from integrating over the first class of two-cycles:

\begin{equation}
\int_{\Delta_A \times T_{mn}} G_4=n_{A,mn} \in \ZZ \ .
\label{genquantization}
\end{equation}
In the GH approximation this integral can be done explicitly and gives  
\begin{equation}
\int_{\Delta_A \times T_{mn}} \Theta^I \wedge J_I =  \, 4 \pi \kappa^I_A \int_{T_{mn}} J_I = \pm \,4 \pi \kappa^I_A L_m\, L_n=n_{A,mn}^{I} \in \ZZ \ ,
\label{eq:quant2}
\end{equation}
where the sign depends on the sign of the two-form $dx^m \wedge dx^n$ inside $J_I$. 
Note that for each $T_{mn}$ there is a single $J_I$ that gives a non-zero contribution, and therefore we get separate quantization conditions for each $I$, whose quantum number we call $n_{A,mn}^{I}$. For a cycle $\Delta_{ij}$ connecting two D6 centers $i$ and $j$, \eqref{eq:quant2} gives
\begin{equation}
	n^I_{ij,mn}=n^I_{i,mn}-n^I_{j,mn}\in \mathbb{Z}\,, \qquad n^I_{i,mn}\equiv 4\pi k_i^I L_mL_n \,,
\end{equation}
whereas for a cycle $\Delta_{ij'}$ connecting the D6 center $i$ to the image of the D6 center $j$, the orientifold action gives $k_j^{I\prime}=-k_j^I$, and hence
\begin{equation}
	n^I_{ij',mn}=n^I_{i,mn}+n^I_{j,mn}\in \mathbb{Z} \,.
\end{equation}
Therefore, flux quantization on the D6-D6 and D6-image-D6 cycles requires both sums and differences of the $n^I_{i,mn}$ to be integer. Equivalently, all $n^I_{i,mn}$ are simultaneously integers or simultaneously half-integers:

\begin{equation} 
	n_{i,mn}^{I}= 4\pi k_i^{I} \, (L_nL_m)\,\ 
		\in \  \begin{cases} \ \mathbb Z \ \quad \ \, \text{for all } i \\
	\mathbb Z+\frac12\ \text{for all }i \end{cases} \,
\label{quantki}
\end{equation}
In addition, we should require flux quantization along the cycles whose Poincar\'e duals, given in  \eqref{swcycle}, are in the chiral spinor-weight class. This requires
\begin{equation}
	\frac{1}{2}\sum_{i=1}^{16}\epsilon_i n_{i,mn}^{I}\in\mathbb Z \,, \qquad \epsilon_i=\pm1 \,, \qquad \prod_{i=1}^{16}\epsilon_i=+1 \,.
	\label{quantization2}
\end{equation}
Together these conditions indicate that for any $I$ and $mn$, the flux vector belongs to the $Spin(32)/\mathbb{Z}_2$ weight lattice
\begin{equation}
	\left(n^I_{1,mn},\ldots,n^I_{16,mn}\right) \in  \left\{n_{i,mn}^{I} \in \mathbb Z\,, \,\,\, \sum_i n_{i,mn}^{I} \in 2\mathbb{Z}\right\} \cup \left\{n_{i,mn}^{I}  \in \mathbb{Z}+\frac{1}{2} \,, \,\,\, \sum_i n_{i,mn}^{I}\in 2\mathbb{Z} \right\} \,.
\end{equation}
The tadpole can be easily written in terms of these quantized numbers:
\begin{equation}
	Q_{\text{Tadpole}} = \frac{1}{2} \sum_{i=1}^{16}\left(n_{i,57}^{4} n_{i,68}^{4} - n_{i,58}^{5} n_{i,67}^{5} -n_{i,56}^{1} n_{i,78}^{2} \right) \,.
\label{eq:tadpoleni}
\end{equation}
This expression is valid regardless of whether the four-form flux is anti-self-dual or not (or equivalently whether three-dimensional Lorentz invariance is imposed). For the Lorentz-invariant solutions, however, the paired flux vectors appearing in \eqref{eq:tadpoleni} are related by the anti-self-duality condition, so that the tadpole is positive. The minimal non-zero tadpole is obtained by turning on only one independent flux type (either $I=4$, $I=5$, or the paired $I=1,2$ contribution), and by choosing the corresponding 16-component flux vector to have the smallest allowed non-zero value of $\sum_i (n_i^I)^2$ in the $Spin(32)/\mathbb{Z}_2$ weight lattice, namely $\sum_i (n_i^I)^2=2$. This gives 
\begin{equation}
	Q_{\text{Tadpole}}^{\text{min}}=1 \ .
\end{equation}

We show in the main text that for this choice one stabilizes the ratio $(L_mL_n)/(L_p L_q)$ (with $m,n$ and $p,q$ determined by which direction $I$ has a non-zero flux). Stabilizing the three moduli requires three independent fluxes $n^4, n^5$ and $n^1=-n^2$ and hence the minimum total tadpole is 
\begin{equation}
	Q_{\text{Tadpole}}^{\text{min}}=3\,.
\end{equation}

\hspace{10mm}

\end{widetext}

\newpage


\bibliographystyle{utphys}      

\bibliography{references}       

\newpage

\end{document}